# Dilution dependence of blocking temperature distribution in the exchange bias system Co$_{1-y}$O(111)/Co(111)

M. R. Ghadimi, M. Fecioru-Morariu, B. Beschoten and G. Güntherodt

II. Physikalisches Institut, RWTH Aachen

Templergraben 55, 52056 Aachen, Germany

In epitaxially grown Co$_{1-y}$O(111)/Co(111) bilayers we have determined the blocking temperature distribution f($T_B$), which is correlated with the antiferromagnetic domain size ($z$) distribution f($z$). The latter is probed by reversing antiferromagnetic domains at successively higher temperatures in the reversed cooling field. The implementation of nonmagnetic defects (y≠0) throughout the antiferromagnet Co$_{1-y}$O is found to give rise to a broadening of the domain size distribution f($z$) within Co$_{1-y}$O as evidenced by a broadened f($T_B$). This broadening is responsible for an enhancement of the exchange bias field.

The exchange coupling at the interface between an antiferromagnet (AFM) and a ferromagnet (FM) below the Néel temperature ($T_N$) of the AFM causes an unidirectional anisotropy in the FM layer, which induces a shift of the hysteresis loop along the magnetic field axis. This phenomenon is called exchange bias (EB).[1,2] For understanding the microscopic origin of EB, the domain state (DS) model was proposed,[3] based on the physics of diluted antiferromagnets in an external magnetic field (DAFF). The DAFF develop into a metastable DS after cooling in an external magnetic field below $T_N$. The domain formation is favoured by intentional dilution, i.e. by implementing nonmagnetic defects in the bulk of the AFM. By reason of statistical distribution of defects in a finite AFM lattice the DS exhibits a distribution f($z$) of AFM domain size ($z$).[3] Each AFM domain carries a local DS magnetization $m_{DS}$, which originates from the uncompensated moments due to the formation of domain walls. Therefore, the total DS magnetization ($M_{DS}$) results from the sum of individual $m_{DS}$. Only the irreversible domain state (IDS) magnetization $M_{IDS}$ gives rise to the EB at the interface to the FM layer. In other words, each AFM domain has its own local unidirectional anisotropy (EB) and its own blocking temperature $T_B$. This depends strongly on the domain size.[4] Hence, within the AFM, f($z$) and therefore the EB field ($B_{EB}$) can be controlled by the number of defects throughout the bulk of the AFM.[3,5-7]

In our previous experimental studies using $Co_{1-y}O(111)/Co(111)$ bilayers we have shown that, besides substitutional defects, $B_{EB}$ can be controlled and increased also by different types of structural defects throughout the volume part of the AFM layer.[7] However, the existence of f($z$) and its dilution dependence was not yet examined. The determination of the blocking temperature distribution f($T_B$), which is correlated with f($z$), yields a qualitative statement about f($z$).[8-10] In this paper, we report on the experimental determination of f($T_B$) and its dilution dependence ($y \neq 0$) in epitaxial untwinned $Co_{1-y}O(111)/Co(111)$ bilayers. The distribution f($T_B$) exhibits for $y \rightarrow 0$ two maxima, one at low temperature and the other one at high temperature near $T_N$. The nonmagnetic defects throughout the antiferromagnet $Co_{1-y}O$ ($y \neq 0$) give rise to a broadening of f($T_B$) and therefore to a broad domain size distribution f($z$) within Co$_{1-y}$O. This broadening is found to be related to an enhancement of $B_{EB}$.

We have studied two samples of 20 nm-Co$_{1-y}$O(111)/10 nm-Co(111)/5 nm-Au, with Co$_{1-y}$O diluted ($y \neq 0$) and undiluted ($y \to 0$) grown by molecular beam epitaxy on MgO(111) substrates. The controlled implementation of nonmagnetic defects at the Co sites of the AFM could be realized by changing the oxygen partial pressure $p(O_2)$ during the growth of the CoO film. The overoxidation of CoO under high $p(O_2)$ yields a Co$^{2+}$-deficient film, denoted as Co$_{1-y}$O, which represents the intentionally diluted sample.[5-7] The epitaxial relationship between MgO, Co$_{1-y}$O and Co was characterized by *ex situ* x-ray diffraction (XRD) using Cu-K$_\alpha$ radiation. The high-angle θ-2θ scans for the diluted (Co$_{1-y}$O grown at $5 \times 10^{-6}$ mbar) and undiluted (CoO grown at $4 \times 10^{-7}$ mbar) samples are shown in Fig. 1. In both samples the Co$_{1-y}$O and Co layers grew with the (111) orientation. Moreover, the XRD patterns show in the case of the diluted Co$_{1-y}$O layer distinct [111] and [333] peaks of the spinel Co$_3$O$_4$. The peak at 44.39° corresponds to fcc Co(111); β denotes the Cu-K$_\beta$ radiation. In addition, *in situ* reflection high-energy electron diffraction (RHEED) was used to characterize the growth of the samples. Further details about the growth of the layer systems used in this study were described previously.[7]

The distribution of $T_B$ was investigated by means of magnetic hysteresis loop measurements using a Quantum Design superconducting quantum interference device (SQUID) magnetometer. The measurements were carried out at 5 K, where thermal activation (TA) of the magnetization within CoO was negligible during the time of the measurement.[11] In order to find the maximum allowable temperature without TA for a certain fraction of the AFM domains, the samples were cooled in the presence of an external magnetic field (field cooling, FC) $B_{cool}$=+0.5 T from 310 K through $T_N$(CoO)=291 K to 5 K. The external field was oriented parallel to the plane of the CoO film along its easy axis [1-21]-direction. At 5 K, we reversed the magnetic field to B=-0.5 T and we measured the first hysteresis loop during 20 minutes immediately after field reversal. The second hysteresis loop was measured after renewed FC from 310 K to 5 K, but after waiting 60 minutes after reversing the magnetic field. The goal was to allow time for the eventual reversal of the AFM domains at that temperature.[11] For the undiluted sample we can observe in Fig. 2 that there were no differences between the hysteresis loops after different waiting times $t_{wait}$. The same behaviour was observed at 100 K as well as for the diluted sample (not shown). This indicates that at least for this time scale of the experiment there were no reversals of the AFM domains and therefore no TA within the CoO(111) layer. This observation is in agreement with our previous measurements of CoO only (without FM layer), which showed that the $M_{DS}$ of the CoO layer remains constant below $T_N$ for a long time, i.e. for a 24 hours measuring time.[12] This gives evidence of an extremely slow thermal relaxation process of CoO. We believe that this is due to the very high anisotropy of CoO.

As a next step we determined $f(T_B)$ by reversing the AFM domains within CoO at different temperatures. For this purpose we did all the measurements at 5 K, first in order to avoid any TA taking place during the time of the hysteresis loop measurement and second in order to determine $f(T_B)$ over a wide range of temperatures. The steps of the procedure (schematically depicted in Fig. 3) are as follows:

**(1)** Set the temperature to 320 K. The CoO layer resides in the paramagnetic phase. Cool the sample from 320 K to 5 K in $B_{cool}$=+0.5 T (FC). In order to reduce the influence of training effect on EB, the field is reversed (B=-0.5 T) at 5 K.[6] Thereby the FM is reversed. Due to initial FC the CoO layer is decomposed into AFM domains of different sizes. The hypothetical domain orientation after FC of the CoO/Co bilayer down to 5 K and after reversing the field is sketched in Fig. 3(a). The orientation of the AFM domains is represented by the direction of the field-cooled uncompensated moments. Due to the high anisotropy of CoO and its grain structure[7] it can be treated as an assembly of independent Ising-type domains. However, these domains are exchange coupled to the FM layer. These have their own local EB and thus own $T_B$, which is strongly domain size dependent.[4] Note, the domain size is determined by the local density of nonmagnetic defects in the bulk of the AFM as well as by the grain size of the CoO layer.[3,7]

(3) Raise the temperature from 5 K to the so-called reversal temperature $T_{rev} > 5$ K, at which some AFM domains reach their $T_B$ with $T_B \leq T_{rev}$ and are "deactivated", i.e. entering the paramagnetic state due to thermal activation [Fig. 3(b)]. Hold the temperature for 60 seconds.

(4) Cool the sample in B=-0.5 T from $T_{rev}$ to 5 K, the so-called reversed FC, and measure the hysteresis loop. The renewed FC "activates" the AFM domains (entering the AFM state), which were previously deactivated at $T_{rev}$. However, they are aligned in the opposite direction [Fig. 3(c)] to the originally set direction [Fig. 3(a)].

This procedure (steps 1-4) is repeated in heating the sample to different $T_{rev}$ [Fig. 3(b)], yielding a "successive" domain reversal within the AFM. In other words, a part of AFM domains within CoO will overcome the energy barriers to reversal and will reverse into the reversed field direction [Fig. 3(c)]. This process depends on the AFM domain size and is correlated with the blocking temperature of the AFM domains.[8-10]. The essential points for the behaviour of the $Co_{1-y}O$/Co bilayers are as follows: First, all measurements were made at 5 K, where TA in CoO can be neglected. Second, the low $T_N$(CoO)=291 K enables the complete reversal of the AFM domains within the CoO layer. Hence, a complete distribution f($T_B$) and thus a complete f(z) of the AFM could be obtained, ranging from 5 K to about $T_N$. The additional advantage of this EB system is the negligible interdiffusion at the interfaces due to the low $T_N$ of CoO.

In order to determine f($T_B$) we extracted the EB field ($B_{EB}$) from the hysteresis loops measured at 5 K. In Fig. 4(a) $B_{EB}$ is shown as a function of the reversal temperature $T_{rev}$, for both the undiluted and diluted samples. $B_{EB}$ starts from negative values for low $T_{rev}$ and increases up to the symmetrical positive value for temperatures near $T_N$ of CoO. In addition, at the same $T_{rev}$ the value of $B_{EB}$ of the diluted sample is generally larger than the one of the undiluted sample. This observation is in agreement with our previous results.[5-7]

The first derivative of $B_{EB}$ with respect to the reversal temperature $T_{rev}$, i.e. $dB_{EB}/dT_{rev}$, can be interpreted as the blocking temperature distribution f($T_B$)[8-10], related to the domain size distribution f($z$). The curves of f($T_B$) of the undiluted and diluted samples are shown in Fig. 4(b). For each sample we observed clearly two maxima of the distributions, one below 75 K and one above 75 K. For undiluted CoO, a plateau is observed in the temperature range from 100 K to 230 K, indicating that no significant reversal of the AFM domains takes place in this temperature range. At higher temperature, close to $T_N$, a narrow distribution can be observed for the undiluted sample. This points out that the undiluted CoO presents a DS with a narrow f($T_B$), i.e. a narrow f($z$). In contrast, the f($T_B$) of the diluted sample shows a broad maximum at high temperature which extends over a much wider range of temperature. Hence, by (intentionally) diluting the AFM CoO, a broad f($T_B$) with a corresponding broad f($z$) is obtained. Due to the static distribution of nonmagnetic defects in the AFM bulk, which hinder the domain-wall motion, the domain walls are pinned at the defects.[3] For the undiluted CoO the domains are scarce, because the formation of domain walls throughout the AFM layer costs much energy.[3] Therefore, we observe a narrow f($T_B$) and thus a narrow f($z$). In an area of the AFM with a definite volume V with large domains, the DS magnetization and therefore $M_{IDS}$ are small.[3,12] This leads to a low $B_{EB}$ compared to the diluted sample as seen in Fig. 4(a). For diluted $Co_{1-y}O$, in which the formation of domains is enhanced, the number of small domains in the same AFM area with the definite volume V is increased.[3] Hence, we observe a broad f($T_B$) and thus a broad f($z$). Therefore, the DS magnetization and thus $M_{IDS}$ are increased and, as a consequence, an increase of $B_{EB}$ results.[12] Moreover, the shift in reversal temperature $T_{rev}$ at which $B_{EB}$ changes its sign [Fig. 4(a)] as well as the shift of the high-temperature maximum in the $dB_{EB}/dT_{rev}$ curves [Fig. 4(b)] with dilution give further evidence for the broadened f($z$) in the diluted sample compared to the undiluted one.

The low temperature peak of f($T_B$) of the diluted sample is higher and wider than the corresponding one of the undiluted sample, but they show the maximum at roughly the same temperature $T_{rev} \approx 20$ K. We believe that this is due to the contribution of isolated AFM spin clusters, the small CoO grains and some defects within these small grains. Moreover, the XRD pattern (Fig. 1) at high oxygen pressure shows for the diluted sample a contribution of grains of the spinel $Co_3O_4$ within CoO. $Co_3O_4$ is also an AFM with $T_N \approx 33$ K and may contribute to the low-temperature peak of f($T_B$) of the diluted sample [Fig. 4(b)]. The contribution of $Co_3O_4$ above $T_{rev}$ exceeding $T_N = 33$ K can be explained by the increase of $T_N$ of $Co_3O_4$ up to 80 K due to exchange coupling with the host CoO.[13]

In conclusion, we have shown that upon diluting the epitaxial AFM CoO by nonmagnetic defects, a wide, double peak blocking temperature distribution f($T_B$) corresponding to a wide AFM domain size distribution f($z$) is obtained. As a consequence, an enhancement of $B_{EB}$ is observed. Additionally, we have shown that no time dependence of the AFM domain reversal is observed for CoO within the time of our measurements.

This work is partially supported by the DFG/SPP1133 and the European Community's Human Potential Program/ NEXBIAS (contract no. HPRN-CT-2002-00269). We would like to thank K. O'Grady for fruitful discussions.

**FIGURE 1**

X-ray diffraction patterns of MgO(111)/Co$_{1-y}$O(111)/Co(111) with Co$_{1-y}$O prepared at $p(O_2)=5\times10^{-6}$ mbar (as diluted) and $p(O_2)=4\times10^{-7}$ mbar (as undiluted).

**FIGURE 2.**

The hysteresis loops of undiluted CoO in MgO(111)/Co$_{1-y}$O(111)/Co(111) taken at 5 K after applying a reversed external field B=-0.5 T at 5 K and after waiting times of 0 min and 60 min.

**FIGURE 3.**

The schematic representation of the antiferromagnetic domains immediately after (a) the FC in $B_{cool}$ and the field reversal ($B_{ext}=-B_{cool}$) at 5 K, (b) the reversed field heating to the respective temperature $T_{rev}$ of the partial AFM domain reversal and (c) the cooling in the reversed field from $T_{rev}$ to 5 K. "inactive" denotes the paramagnetic state of the domains. The orientation of the AFM domains is represented by the direction of the field-cooled uncompensated moments.

**FIGURE 4.**

(a) $B_{EB}$ and (b) distribution profiles of the blocking temperature as function of $T_{rev}$ for undiluted (▲) and diluted (□) samples. At the respective $T_{rev}$ a certain fraction of AFM domains enters the paramagnetic state and is reversed upon cooling to 5 K in the reversed field ($B_{ext}=-B_{cool}$).

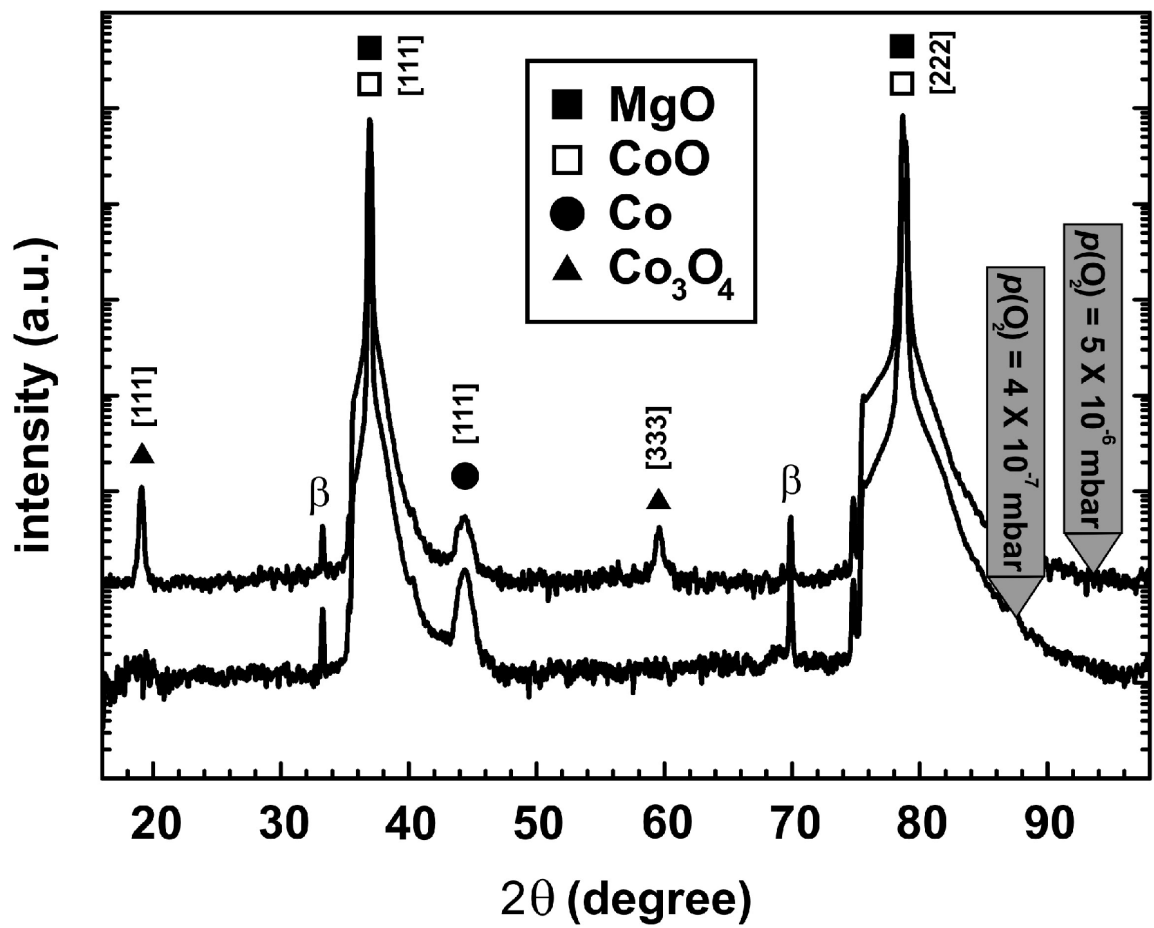

Figure 1: M. R. Ghadimi et al.

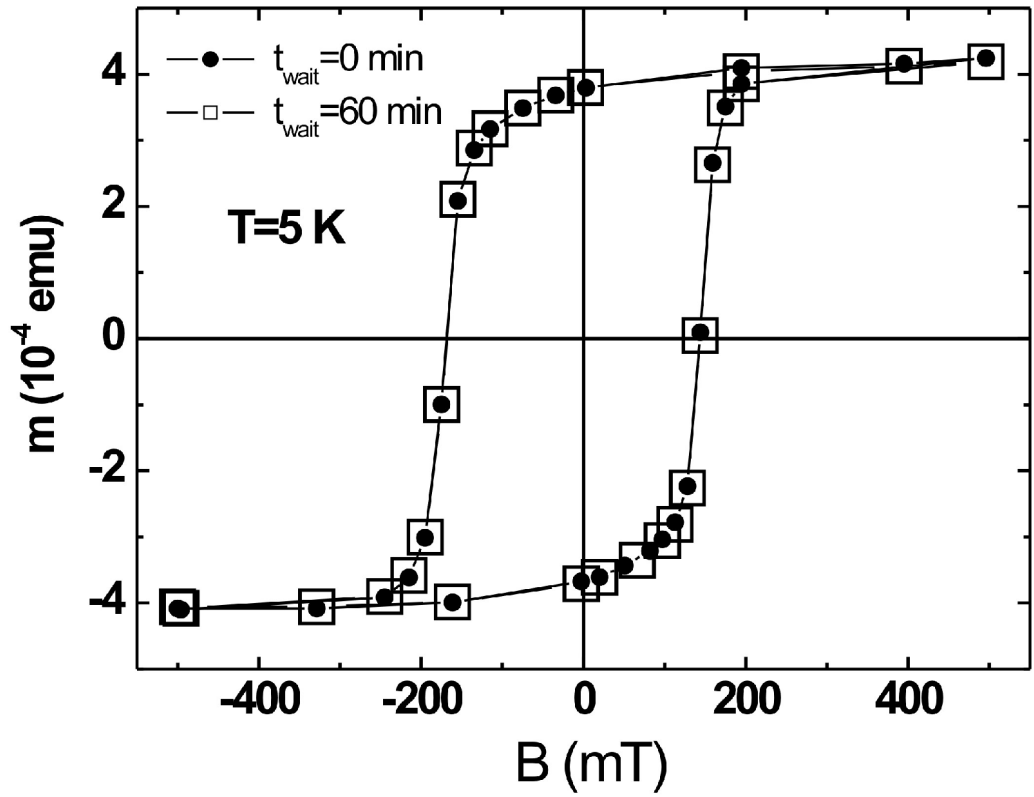

Figure 2: M. R. Ghadimi et al.

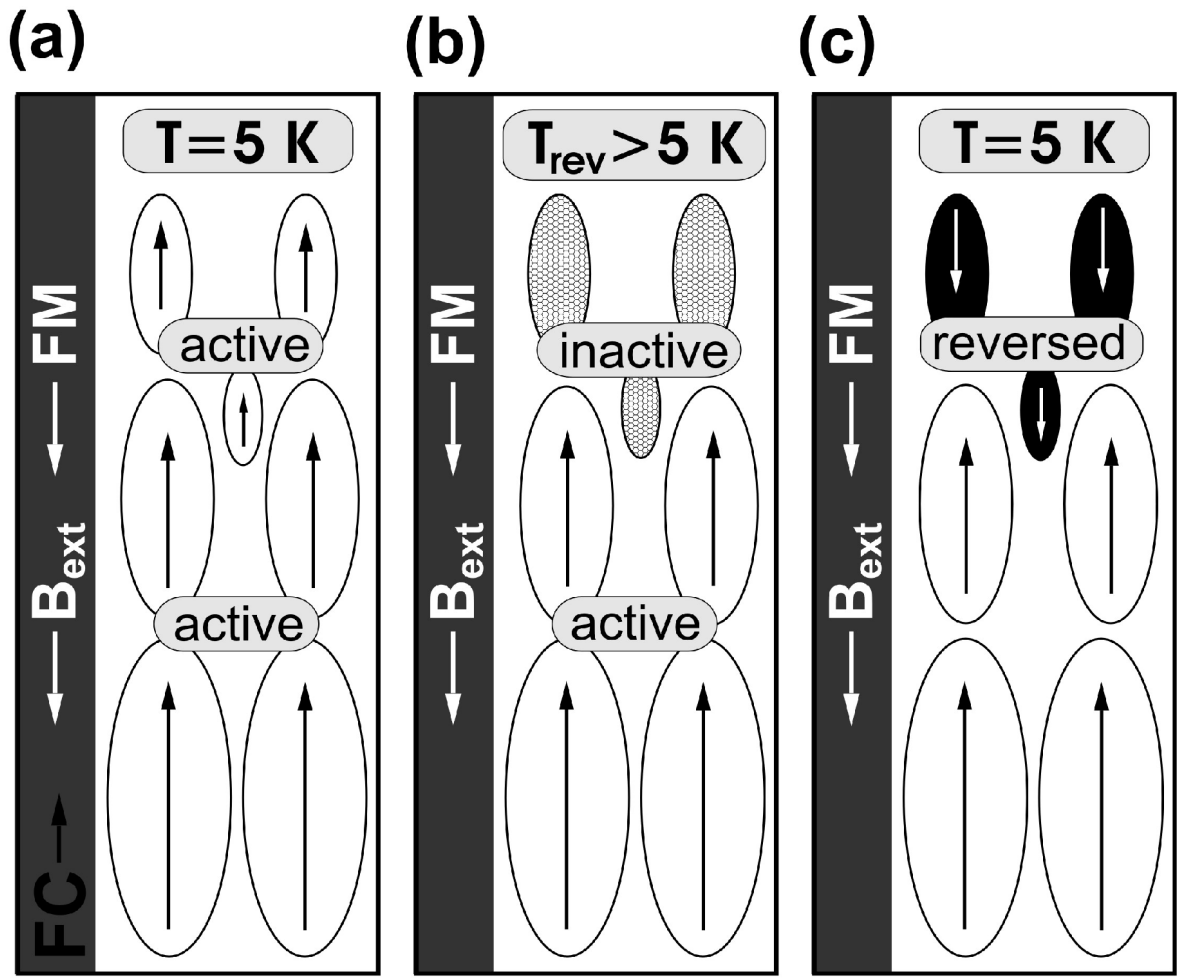

Figure 3: M. R. Ghadimi et al.

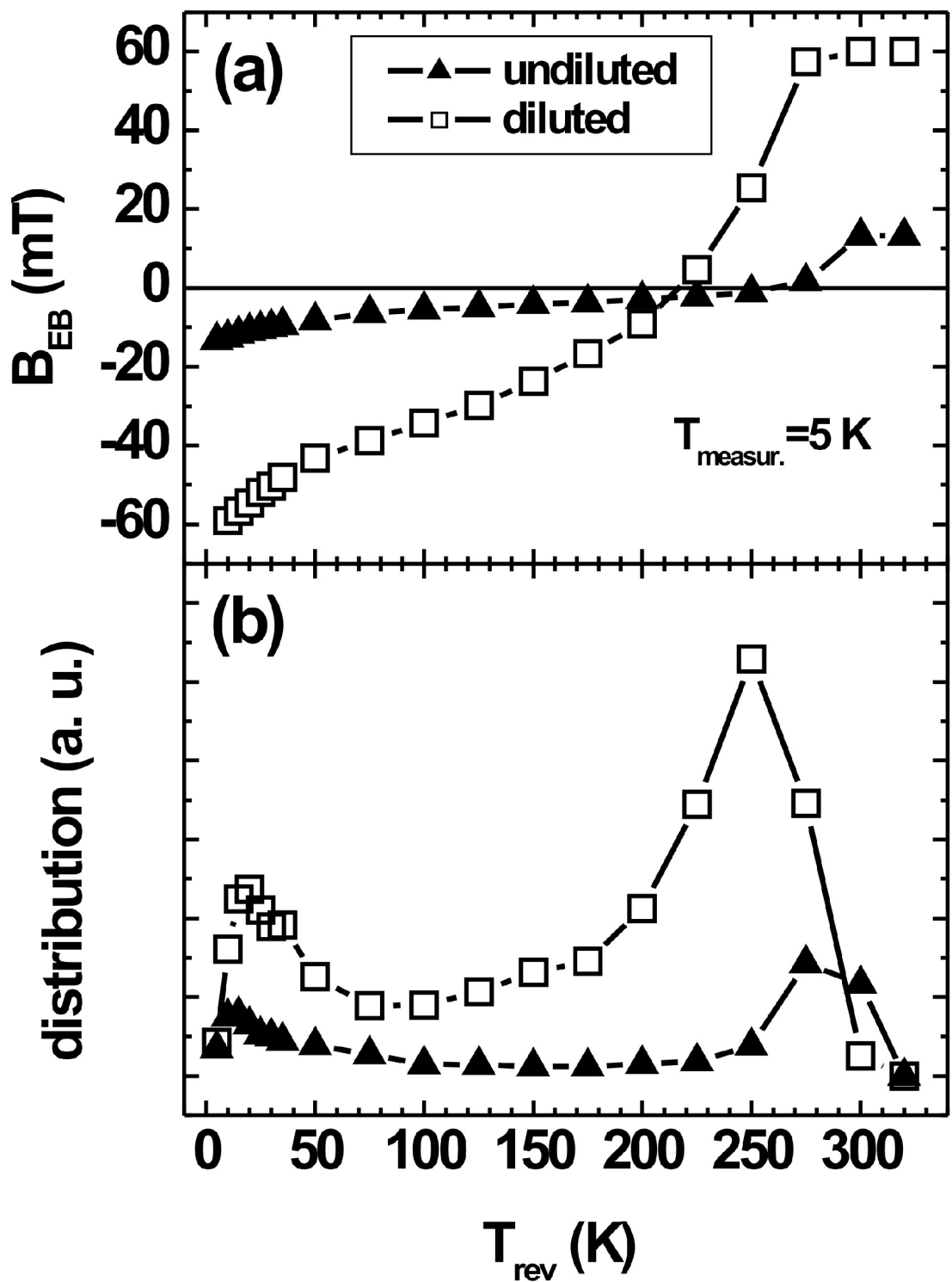

Figure 4: M. R. Ghadimi et al.